\documentclass[twocolumn,showpacs,preprintnumbers,amsmath,amssymb]{revtex4}
\usepackage{graphicx}
\usepackage{dcolumn}
\usepackage{bm}

\begin{document}

\title{Effects of excess or deficiency of oxygen content on the electronic structure of high-$T_C$ cuprates.}
\author{T. Jarlborg$^1$, A. Bianconi$^{2}$, B. Barbiellini$^3$, R.S. Markiewicz$^3$, A. Bansil$^3$}

\affiliation{
$^1$ DPMC, University of Geneva, 24 Quai Ernest-Ansermet, CH-1211 Geneva 4, Switzerland
\\
$^2$ RICMASS, Rome International Center for Materials Science Superstripes, Via dei Sabelli 119A, 00185 Rome, Italy.
\\
$^3$ Department of Physics, Northeastern University, Boston, Massachusetts 02115, USA
}




\maketitle


{\bf Abstract.} 

Band structure calculations are presented for large supercells of Ba$_2$CuO$_4$ (BCO) with O-vacancies in planar or apical positions, and of superoxygenated La$_2$CuO$_4$ (LCO) with oxygen interstitials in the La$_2$O$_2$ layers. It is found that apical oxygen vacancies in BCO act as electron dopants and makes the electronic
structure similar to that of hole doped LCO. Excess oxygen interstitials forming wires in the La$_2$O$_2$ layers of LCO are shown to yield a much larger density-of-states at the Fermi energy than for the stoichiometric compound related with a segmentation of the Fermi surface.
Anti-ferromagnetic (AFM) spin fluctuations are strengthened by O-vacancies in BCO as well as by oxygen interstitials in LCO, but are strongly suppressed in O-deficient LCO.  
Our results indicate the complexity of doping by O-vacancies, and by ordered defects that are a significant factor controlling the electronic properties of cuprates.

\vspace{3mm}


\subsection{Introduction}

High-$T_c$ cuprate superconductors in their normal state show many unusual properties such as pseudogaps, stripe-like
charge/spin modulations with particular energy/doping dependencies, Fermi-surface (FS) "arcs" in
the diagonal direction, 'kinks' and 'waterfalls' in the band dispersions, 
anomalous isotope effect, and phonon softening \cite{dama}-\cite{armi}.
It is generally believed that stripe-like modulations of the Cu-spin arrangements are important for doped cuprates. 
Band results for long '1-dimensional' (1-D)
supercells, calculated by the Linear Muffin-Tin Orbital (LMTO)
method in the local spin-density approximation (LDA), 
show large spin-phonon coupling (SPC)
within the CuO planes \cite{tj1,tj3}. 
The LMTO results have been used to parameterize
the strength of potential modulations coming from
phonon distortions and spin waves of different length \cite{tj5}.
Phonon softening, dynamical stripes, correlation between $\bar{q}$ and $x$,
smearing of the non-diagonal part of the FS, and abrupt disappearance of spin
fluctuations at a certain $T^*$, are all possible consequences of SPC within a 
rather conventional band picture\cite{tj5,tj6}. 
Weak ferromagnetism is a possibility at very high hole doping levels \cite{bj,son}.  
The focus has been for many years on the intrinsic electronic response of the CuO$_2$ plane following electronic 
doping pushing the chemical potential away from half filling. The details of the structure of spacer layers 
intercalated between the CuO$_2$ planes and the role of dopants, atomic substitutions, defects or oxygen 
interstitials inserted there to convert the undoped antiferromagnetic CuO$_2$ plane into a superconductor 
has not been much considered so far.

Recently there is a growing interest to control the superconducting  $T_c$ by changing the structure of the 
spacer layers. It is known that maximum superconducting  $T_c$  at optimum doping increases 
from 20 to 130 K by manipulation of the spacer layers structure \cite{alloys} and the spatial distribution 
of defects inserted there. 
For example the maximum superconducting  $T_c$  in La$_2$CuO$_4$ (LCO) achieved through La/Sr or La/Ba exchange 
is 35 K while with the maximum $T_c$ in superoxygenated La$_2$CuO$_{4+\delta}$ can reach 45 K.  
Recently the ordering of oxygen interstitials has been shown to control the superconducting  $T_c$   
\cite{frat,pocc}. High $T_c$'s have been reported for oxygen deficient Sr$_2$CuO$_{4-\delta}$ 
\cite{liu,gao,geba} and in other cuprates with ordered defects \cite{chma}.  Moreover the dopants and their 
self organization have been shown to control the critical temperature in pnictides \cite{garcia12} 
and chalcogenides \cite{ricci11}.

Here we present aspects of our results of band structure calculations on supercells of Ba$_2$CuO$_4$ (BCO) 
or La$_2$CuO$_4$ (LCO) with various distributions of O-defects for the purpose of gaining insight into how 
the electronic structure and magnetic interactions are modified by the presence of such defects. 
For oxygen deficient structures it matters if the vacancies
are located within the CuO$_2$ planes or in apical positions.
In the latter case the vacancies act as electron dopants of
the Cu-d band, so that BCO with a substantial apical-
vacancy concentration would approach the band filling
of hole doped LCO. Excess oxygen interstitials inserted in the rocksalt La$_2$O$_2$ layers of
LCO occupy the oxygen interstitial positions as oxygen ions in fluorite Nd$_2$O$_2$ spacer layer of Nd$_2$CuO$_4$.

We show here the breakdown of the common rigid band model for doped CuO$_2$ planes.  Defects induce a band folding 
and opening of partial gaps at specific points in the k-space and produce multiple mini-bands crossing the 
Fermi level as it has been proposed \cite{per}. Moreover the ordered oxygen interstitials yield a much larger 
density-of-states at the Fermi energy than for the stoichiometric compound, favoring enhanced spin fluctuations.

A more complete discussion of these results will be undertaken elsewhere \cite{sco,bianc}.

\subsection{Method of calculation}

Ab-initio LDA-LMTO band calculations are carried out for
supercells of R$_2$CuO$_{4\pm\delta}$, with R=La or 
Ba (LCO or BCO). In the case of
oxygen vacancies in LCO and BCO, the calculations are based on supercells with a total of 112 sites
\cite{sco}. The supercell is obtained via a 2x2x2 extension of the 
basic anti-ferromagnetic (AFM) cell, R$_{32}$Cu$_{16}$O$_{64-n_V}$, where $n_V$ is the number of
O-vacancies in the cell. Different concentrations $\delta=n_V/16$ of vacancies either in the planes (P) or in apical (A)
positions are realized through a randomized (not clustered) distribution of vacancies in
R$_2$CuO$_{4-\delta}$.

For simulating ordered stripe-like interstitial oxygens, we
first insert empty spheres in all interstitial positions at $(\frac{1}{2},0,\frac{1}{2}c)$ and 
$(0,\frac{1}{2},\frac{1}{2}c)$, and then occupy one or two of these positions with additional
oxygen interstitials. In total, this yields a supercell with 72 sites for a 4x1x1 extension of the basic AFM cell
along the diagonal [1,1,0]-direction,
La$_{16}$Cu$_8$O$_{32+N}$, where $N$ is the number of additional oxygens.
There are one or two adjacent rows of additional O in the supercells \cite{bianc}.
The atomic spheres are reduced compared to
the calculations for the 112-site supercell, but the results for defect free LCO are very similar
in the two sets of calculations. 

Spin-polarized calculations are made with applied magnetic fields on selected Cu-sites in order to generate AFM order. Other details of our band calculation methods have been published elsewhere \cite{apl,sco,bianc}. It should be noted that a rigid band picture \cite{bansil1} or its variants \cite{bansil2} have often been invoked in describing the doping evolution of the overdoped and optimally doped cuprates \cite{bansil3,bansil3a}. Our large supercell treatment here goes beyond the simple rigid band model or possible mean-field type approaches \cite{bansil4} to elucidate the local electronic and magnetic properties of the system.

\begin{figure}
\includegraphics[height=7.0cm,width=8.0cm]{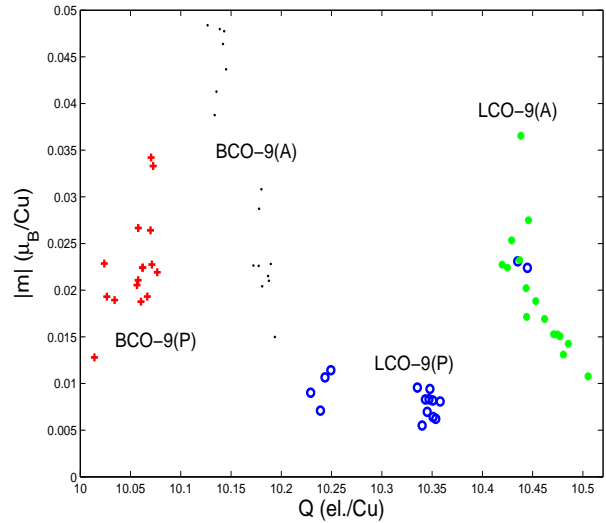}
\caption{Absolute value of the magnetic moment $m$ as function of the valence charge
for each Cu-atom in BCO and LCO with 9 apical or planar O-vacancies.}
\label{figqm}
\end{figure}

\begin{table}[b]
\caption{\label{table1} Valence charge $Q_{Cu}$ (in electrons/per Cu), the change $\Delta Q_{Cu}$ with respect to the 
undoped case, and the magnetic moment $m$ (in $\mu_B$ per Cu) as a function of the number $(n_V)$
of planar (P) or apical (A) oxygen vacancies 
in BCO and LCO. A magnetic field of $\pm$2.5 mRy was applied to generate AFM order on Cu.}
\vskip 5mm
\begin{center}
\begin{tabular}{l c c c c c c}
  \hline
     &  & BCO &  &  &~~ LCO &  \\
  \hline
 $n_V$ &~~ $Q_{Cu}$ & $\Delta Q_{Cu}$ & $\bar{m}$ &~~~ $Q_{Cu}$ & $\Delta Q_{Cu}$ & $\bar{m}$ \\
  \hline \hline
  9(P)  &~~ 10.055 & -0.010 & 0.024 &~~~ 10.332 & -0.063 & 0.010 \\
  8(P)  &~~ 10.056 & -0.009 & 0.022 &~~~ - & - & - \\
  5(P)  &~~ 10.057 & -0.007 & 0.024 &~~~ 10.368 & -0.026 & 0.025 \\
  1(P)  &~~ 10.062 & -0.002 & 0.021 &~~~ 10.395 & -0.000 & 0.073 \\
  0 &~~ 10.064 &  0.000 & 0.020 &~~~ 10.394 & 0.000 & 0.069 \\
  1(A) &~~ 10.071 &  0.007&  0.021 &~~~ 10.420 & 0.026 & 0.069 \\
  5(A) &~~ 10.107 &  0.043 & 0.028 &~~~ 10.450 & 0.056 & 0.024 \\
  8(A) &~~ 10.145 &  0.081 & 0.038 &~~~ - & - & - \\
  9(A) &~~ 10.163 &  0.099 & 0.035 &~~~ 10.456 & 0.062 & 0.022 \\
  \hline
\end{tabular}
\end{center}
\end{table}

\subsection{Results for O-deficient BCO and LCO}

Unpolarized calculations show that apical or planar oxygen vacancies act very differently with respect to 
d-band filling and charge on Cu \cite{sco}. Apical vacancies in BCO lead to a charge transfer 
toward Cu and an upward shift of $E_F$ relative to the d-band, so that the band structure of BCO
with many vacancies can approach that of hole doped LCO \cite{note1}. Apical vacancies in LCO
have a similar but weaker doping effect, whereby LCO becomes electron doped.
Vacancies in the planes of LCO lead to hole doping, while in BCO the effect on doping is small. 
These results are interesting in view of recent reports of high $T_C$ in oxygen deficient
SCO \cite{geba}, although possibility of some type of ordering has been implicated \cite{chma,gao}.
A question that arises naturally in this connection is whether the propensity for AFM fluctuations 
correlates with the effective doping level, since spin fluctuations are probably involved in the mechanism of
superconducting pairing. Accordingly, in our calculations we mimic an AFM order by applying positive or negative 
magnetic fields on every
second Cu. We have taken the AFM structure and the field amplitudes
to be identical in all cases investigated in order to make direct comparisons between BCO and LCO with different
types of defects.

Our main results are summarized in Table I. For vacancies in CuO planes of BCO there is, in addition to the weak
hole doping within the Cu-d band, a tendency toward stronger AFM fluctuations. The enhancement is
similar for vacancies on the apical positions, in spite of the clear
electron doping in this case. The evolution of AFM fluctuations for O-vacancies in LCO is very different.
Undoped LCO has strong enhancement of AFM waves that leads to a stable AFM configuration
and an insulator. Well converged LSDA calculations do not find this state, but it has been
shown that LMTO calculations with off-center linearization energies of the Cu-d states lead
to slight localization and stable AFM order \cite{tj4}.  Both apical and planar vacancies quickly suppress
the tendency for anti-ferromagnetism as seen in Table I.  This agrees qualitatively with early calculations
for a single O vacancy in small unit cells of LCO \cite{ster}.

Interestingly, we find that the electron doping effect in BCO with apical vacancies is strong,
and one may ask if the properties of such crystals are similar to those of hole doped LCO.
Our results indicate that it is not exactly so, since anti-ferromagnetism in BCO with the highest electron
doping is stronger than in LCO where maximal hole doping is made via planar O-vacancies. In LCO, 
it seems that breaking the near neighbor order through apical or planar O-vacancies will effectively
quench AFM fluctuations, much like the effect of Sr doping. An
example can be seen in 
Fig. \ref{figqm}, where the local moment on Cu is plotted as a function of the local Cu valence charge
for configurations with 9 O-vacancies. These results suggest that local order is probably
an important factor for anti-ferromagnetism.
The increasing strength of spin fluctuations with increasing number of apical vacancies 
provides a possible explanation for why oxygen deficient SRO
can have a higher $T_C$ than the traditional hole doped LCO. Doping is usually thought to be determined by
the La vs. Sr/Ba substitution, but our results show that oxygen deficiency may
have large effects on doping as well. For instance, electron doping usually carried out through La/Nd-Ce
substitution is accompanied with a structural change of the apical O position, but our results show that apical 
O-vacancies in LCO could provide an alternate route for electron doping keeping the structure intact. 

\begin{figure}
\includegraphics[height=7.0cm,width=8.0cm]{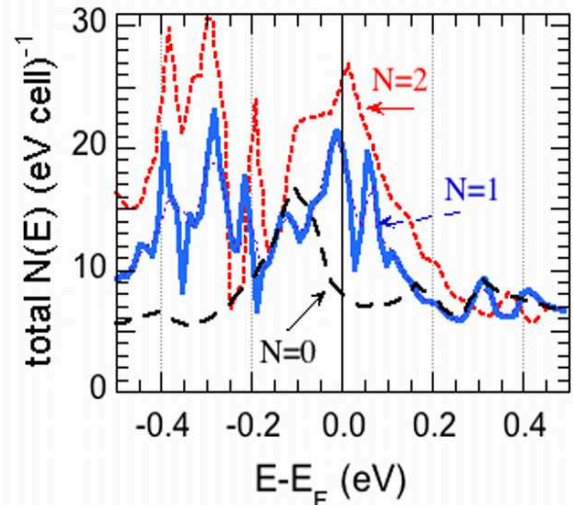}
\caption{The total DOS of the undoped  La$_{16}$Cu$_8$O$_{32}$
near the Fermi level is compared with the DOS of doped 
superoxygenated La2CuO4 with additional N=1 and N=2 oxygen interstitial in the supercell.}
\label{figi3}
\end{figure}

\begin{figure}
\includegraphics[height=7.0cm,width=8.0cm]{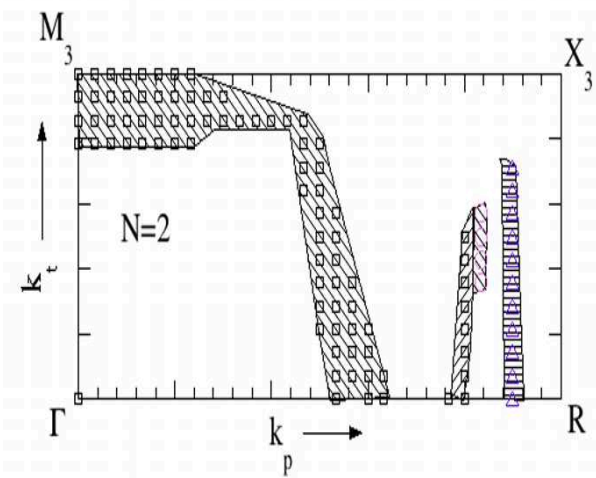}
\caption{The calculated FS for La$_{16}$Cu$_8$O$_{32+N}$ for $N=2$.
This FS can be understood from a folding of the original circular FS into 
the Brillouin Zone of the supercell (given by $\Gamma-R-X_3-M_3$), but
the partial gaps appears only for N$>$0.}
\label{figi1}
\end{figure}

\subsection{Results for stripes of oxygen interstitials in the spacer layers}

Rows of single or double interstitial O's oriented along [1,1,0] make the DOS at $E_F$ larger 
than in undoped LCO by factors of 2.5 or 3, respectively, see Fig \ref{figi3}. 
The new O-p bands associated with the interstitials
are found to lie a few eV below $E_F$, but hybridization effects make the bands at $E_F$ narrower.
The interstitial oxygens and the neighboring
atoms acquire most of the additional DOS. This increase in DOS comes about because the band
at $E_F$ is broken up by the new periodicity with smaller dispersion between the resulting small gaps. A folding
of the original almost circular LCO Fermi surface (FS) into the new Brillouin Zone (BZ), given by $\Gamma - R - X_3 - M_3$,
produces a FS similar to the one in Fig. \ref{figi1}.
The computed FS for the undoped supercell (N=0) of LCO agrees even more 
with the model where circular FS in the original BZ has been downfoled into the BZ of the supercell. One difference is
that the FS's for the supercells containing extra O are segmented with gaps between the branches. 
The three branches in Fig. \ref{figi1}
corresponds to different sections of the circular FS in the original BZ. Furthermore, the calculated
band mass increases when one or two O atoms are added, and bands become gapped near 
the boundaries of the new BZ.
A general upward displacement of the FS branch in the left part of Fig. \ref{figi1} can be understood from an increased
radius of the original FS. This represents hole doping, which is consistent with
effective charges on Cu: Addition of each interstitial O in the cell removes
about 0.04 electrons from each Cu.

The new periodicity sets up a potential modulation along the unit cell and as expected induces weak pseudogaps in the DOS near $E_F$. It turns out that $E_F$ falls on a DOS peak above
the gap energy, both for one and two extra oxygen atoms. Thus, the observation of ordering of oxygen interstitials \cite{pocc,poccia2} might be a promising way for creating potential modulations, pseudogaps and enhanced $N(E_F)$ 
for boosting superconductivity \cite{apl}. Furthermore, FS segmentation and
small gaps separating mini bands could provide favorable conditions for Lifshitz transitions in connection with 
other mechanisms for enhanced $T_C$ \cite{inno}. Notably, the enhanced $N(E_F)$ favors AFM fluctuations in general. The calculation of modulated AFM spin waves 
shows a moderate increase of the exchange enhancement when one or two O-interstitials are added, but the
effect concerns mostly Cu sites far from the interstitials. The strong hybridization with the additional
O-p electrons, which are not magnetically active, is probably the cause of the limited effect on nearby
Cu sites. 

\subsection{Conclusion}

Our study shows that electronic properties of cuprates can be modified substantially by the presence of oxygen defects. 
In particular, vacancies among apical and planar oxygen sites lead to very
different charge filling within the Cu-d band, and strong changes in the AFM couplings in the system. Moreover, AFM fluctuations in BCO and LCO behave quite differently as a function
of doping. The effective doping depends not only on La/Ba substitutions, but also on the specific nature of oxygen vacancies. For excess O, we follow the suggestions from experiments and assume that
interstitial oxygens order into stripes. The new periodicity makes the DOS high
at $E_F$, which leads to some enhancement of AFM fluctuations. The FS branches remain fairly intact and identifiable for different doping, but mini-gaps are formed to make the FS segmented. Heavier band masses are 
evident as well as a slight hole doping for increasing O content. These changes are normally favorable to a higher $T_C$.  

\subsection{Acknowledgements}

This work is supported by the U.S. Department of Energy contract number DE-FG02-07ER46352, and benefited from the allocation of supercomputer time at the NERSC and Northeastern University's Advanced Scientific Computation Center (ASCC).


\begin{thebibliography}{10}
\bibitem{note1} Note that Ba and Sr have one less electron than La, which makes BCO and SRO, Sr$_2$CuO$_4$, an extremely hole doped version of LCO.
\end{thebibliography}
\end{document}